\documentstyle[psfig,conf_iap]{article}

\def\lesssim{\mathrel{\hbox{\rlap{\hbox{\lower4pt\hbox{$\sim$}}}\hbox{$<$}}}}
\def\gtrsim{\mathrel{\hbox{\rlap{\hbox{\lower4pt\hbox{$\sim$}}}\hbox{$>$}}}}
\newcommand{\eg}{{\it e.g.}}
\newcommand{\sqrdeg}{\Box^{\circ}}
\newcommand{\etal}{{\it et al.}}
\begin{document}
\heading{%
%
The Impact of AXAF \& XMM on Measurements of $\Omega_0$ from Cluster Abundances
%
} 
\par\medskip\noindent
\author{%
A. Kathy Romer $^{1}$
}
\address{%
Physics Department, Carnegie Mellon University, Pittsburgh, PA15213, USA.
}
 
\begin{abstract}
Two major X-ray satellites, AXAF \& XMM, will be launched in 1998/1999. 
Both satellites have spectral and imaging capabilities,
and are several times more sensitive than either ROSAT or ASCA.
AXAF \& XMM will yield significant scientific returns in many fields
of astrophysics. We concentrate here on how AXAF \& XMM will improve 
constraints on the density parameter, $\Omega_0$, via measurements
of the cluster abundance at high redshift. 
\end{abstract}

\addtocounter{footnote}{1}

\section{Biasing $\Omega_0$ Using Observed Cluster Abundances}

In a high $\Omega_0$ universe, the density of the most massive 
clusters will be vanishingly small at high redshift.
This density rises rapidly as $\Omega_0$ decreases, meaning that the 
discovery of only one or two massive clusters at a redshift of 
$z$$\sim$1 has the potential to rule out $\Omega_0$$=$1 to high significance \cite{bahcall}. This sensitivity has meant
that measurements of cluster abundances\footnote{By ``abundances'' 
we mean the number density of clusters as a function of mass and $z$.} have become a popular means by which to
constrain $\Omega_0$. However, one has to be extremely cautious when
making cluster abundance measurements, since any observational bias which mimics an 
under (or over) abundance of massive clusters will lead to an over 
(or under) estimate of $\Omega_0$. 

For example, one could mimic an under abundance of massive clusters in an 
X-ray selected survey, such as the EMSS \cite{gioia90} \cite{henry92}, 
if the optical follow-up was incomplete: Optical follow-up becomes 
increasingly difficult as redshift increases and so there is a finite probability 
that some high $z$ clusters will be missed and that other clusters will be assigned
incorrect redshifts. The optical follow-up of the EMSS cluster 
sample is now complete \cite{gioia94}, but in an earlier version of the 
catalog \cite{henry92}, one high redshift ($z$$>$0.5) cluster was missed 
(MS1610.4) and two more had underestimated redshifts (MS1054.4 \& 
MS1137.5). Small errors like these have a significant effect on
measured parameters, such as the significance of measured luminosity
evolution \cite{nichol97} and on the value of $\Omega_0$ \cite{reichart}. 

To date, the analysis of cluster abundances has been made in the 
context of the Press-Schechter formalism, which gives 
an analytical relation for the number density of {\em virialized} 
dark matter halos as a function of mass and redshift\footnote{This formulism has
been shown to be very accurate \cite{lacey93} but will, no doubt,
be replaced in coming years as larger and more accurate n-body simulations
become available \cite{colberg}.} \cite{ps74}. When even a small fraction 
of the high redshift clusters in a given catalog have
artificially high masses, then the Press-Schechter derived value of $\Omega_0$ will 
be too low. This situation could easily occur in an X-ray 
selected survey, if one has to rely on {\it global} X-ray luminosity or 
temperature measurements to derive the cluster masses: If a cluster 
has significant subclustering, \eg~MS1054.4 \cite{donahue}, then the derived mass can be higher than the virial mass.

Viana \& Liddle conclude that, ``at present the observational data 
and the theoretical modelling carry sufficiently large associated 
uncertainties to prevent an unambiguous determination of $\Omega_0$'' \cite{viana98}.
This is illustrated by the fact that the EMSS yields best fit
values that range from $\Omega_0$=0.2 \cite{bahcall} to 
$\Omega_0$=1 \cite{blanchard}.

\section{The Impact of AXAF and XMM}

The two main observational issues that need to be addressed when 
attempting to measure $\Omega_0$ from cluster abundances are; ({\it i})
the completeness of the cluster catalog under study and ({\it ii}) the accuracy 
of the cluster mass estimates. We describe the positive 
impact that AXAF \& XMM will have in both areas below.

\paragraph{Improving cluster mass estimates:}

It has been shown that there is a tight relationship between cluster 
mass and X-ray temperature ($T_x$) in a virialized system \cite{evrard96}. 
Existing $T_x$ data, derived from ASCA and GINGA observations \cite{henry97}, 
provide only weak constraints on $\Omega_0$ \cite{viana98}, 
but we can expect these constraints to tighten dramatically after the 
launch of AXAF and XMM. These satellites will provide $T_x$ values 
more accurately, and more efficiently, than ever before. 
As an  illustration, let us compare the  expected countrates\footnote{These 
countrates were derived using HEASARC 
W3PIMMS webpage and assuming an 8keV Raymond-Smith spectrum and the 
total [0.3-3.5 keV] flux quoted in \cite{gioia94}.} for the most
luminous cluster in the EMSS (MS0015.9, $z$=0.54) in the AXAF ACIS-I 
camera (0.15 s$^{-1}$), the XMM EPIC pn-camera (0.57 s$^{-1}$) and 
an ASCA SIS camera (0.05 s$^{-1}$). With $\simeq$2000 photons, or a 13 
ks ACIS-I observation, one can measure $T_x$ for this cluster
to a reasonable accuracy ($\delta T_x/T_x$$<$0.2) \cite{henry97}. 

It is likely, given the excellent spatial resolution of AXAF 
\& XMM ($0.5''$ and $10''$ respectively) that many guest observers
will request exposure times long enough to make spatially resolved 
temperature maps of high redshift clusters. (With ASCA, these sorts 
of maps have only been feasible for high flux, low redshift, 
clusters \cite{mark98}.) These efforts should be encouraged, since
temperature maps allow one to correct $T_x$ for the influence of 
shock fronts at subcluster boundaries and of cooling flows in the
cluster core. The exposure times required will be high, but over 
the lifetime of the satellites ($\sim10$ years), we can expect
temperature maps to become available for a significant fraction 
of known X-ray clusters at $z>0.3$. For example, to measure a 
$T_x$ value in 4 independent radial apertures for MS0015.9, one 
would require $\simeq50$ ks with AXAF or $\simeq14$ ks with XMM.

\paragraph{Towards new cluster samples:}

In addition to the EMSS, which was produced from Einstein-IPC data, 
there are now several samples of X-ray selected, high redshift, 
clusters based on ROSAT PSPC data \eg ~\cite{burke}\cite{romer}\cite{viklinin}. 
These ROSAT surveys cover smaller areas\footnote{ROSAT-PSPC surveys based on
pointing data will never cover more than $\sim 200\,\sqrdeg$ because the PSPC instrument 
was retired after only 4 years in service.} ($17-200\,\sqrdeg$) than the EMSS ($40-730\,\sqrdeg$), which is a distinct disadvantage, since it is 
areal coverage, not flux limit, that determines the number of high $z$, 
high mass, clusters in a given survey. For example, in only a 1 ks XMM observation
one could detect a massive cluster at $z$$=$1 to a signal-to-noise 
greater than 10. (Here we define a ``massive cluster'' to be one with a 
luminosity greater than $L_{\star}$, where $L_{\star}$$\simeq$3e44 erg/s \cite{ebeling}.) 
But, since these clusters are so rare beyond $z$$=$0.3, one would need to make 
$\simeq$420, non-overlapping, XMM pointings to guarantee a single detection. (This
estimate was based on the number of 0.3$<$$z$$<$1, $L_x$$>$$L_{\star}$ clusters in 
the EMSS \cite{gioia94}.) 

Apart from a possible ``XMM Slew Survey'', which would have an effective 
exposure time of less than $100$ seconds (David Lumb, private 
communication), there are no plans to use either XMM or AXAF as survey instruments.
Any new cluster catalogs would, therefore, have to be based 
on serendipitous detections. Given the growing number of ROSAT 
derived cluster catalogs, and the huge areal coverage of the EMSS, 
would yet another serendipitous cluster survey be worthwhile? 
We suggest that is it not only worthwhile, but imperative. This is
because all surveys to date have been based on less than ideal X-ray
data, meaning one cannot fully define their selection functions, 
and hence completeness, using simulations. For example, 
({\it i}) several of the EMSS clusters were detected at
only the $5\sigma$ level \cite{nichol97} and ({\it ii}) ROSAT-PSPC cluster
surveys which rely on source extent have problems with blended emission 
(blends make up $\sim 50\%$ of the cluster candidates in the Bright SHARC 
sample \cite{romer}). The only way to remove observational biases like
these is to create a new cluster sample based on  higher quality X-ray 
data.
 
If the XMM-EPIC camera remains in service during the whole lifetime
of the satellite, then one could use it to build up an X-ray cluster
survey with the same areal coverage as the EMSS. Under the conservative 
assumption of only 3 pointings per day, then XMM would
be able to cover a total area of 2200 $\sqrdeg$ in 10 years.  
(The XMM field of view is $0.2\,\sqrdeg$ compared to $1.6\,\sqrdeg$ for
the Einstein-IPC.) Not all of the available area will be of use to a 
cluster survey, however, since some will fall in the galactic plane and some 
will be in the field of diffuse X-ray sources, such as low $z$ 
clusters and supernova remnants. Using the ROSAT archive as 
a guide, one can expect that only $\simeq$1/3 of the XMM pointings
will be suitable for a serendipitous cluster survey, but this would
still yield $\gtrsim 700\,\sqrdeg$ over the lifetime of the satellite\footnote{It 
will not be practical to carry out a serendipitous cluster survey with AXAF
because its FOV for imaging is even smaller ($0.08\,\sqrdeg$) than that
of XMM. In addition, unlike XMM, AXAF is not able to produce imaging 
data when the diffraction gratings are in place.}. Such a survey will
be aided both by the planned pipeline processing at the XMM Survey Science Center 
and by the spatial resolution of the EPIC cameras: The XMM spatial resolution 
is better, even at the edges of the FOV, than the on-axis resolution of the 
ROSAT-PSPC. This means that far fewer blended point sources will be falsely 
flagged as cluster candidates, which in turn eases optical follow-up.
In addition, a new XMM cluster sample will require much less X-ray follow-up
than the EMSS or the ROSAT-PSPC samples. This is because the majority 
of XMM exposures are expected to be at least ten times longer than the 
1 ks required for a  $10\sigma$ detection of a $z$=1, $L_{\star}$ cluster. 
Therefore, most serendipitous cluster observations will yield sufficient 
counts to allow cluster profiles and global $T_x$ values to be measured 
directly. 

\section{Summary}

Measurements of cluster abundances allow one to place strict
constraints on the value of $\Omega_0$ which are independent 
from those derived from high redshift supernovae or from Cosmic 
Microwave Background fluctuations. After the launch of AXAF \& XMM,
it will be possible to remove several observational biases that have
dogged previous attempts to measure $\Omega_0$ from cluster
abundances. Progress will come via the derivation of accurate 
virial temperatures for a large number of clusters and via the 
development of a new, large area, cluster survey. 

\acknowledgements{The AXAF/XMM predictions described here were funded 
in part by NASA grant NAGW3288. We acknowledge the IAP 
for providing financial support for travel to the conference and RCN/ACR
for encouraging remarks.}

\vfill
\end{document}